\documentclass[10pt,preprint2]{aastex}

\newcommand{\Msun}{\hbox{$\hbox{M}_\odot$}}
\newcommand{\Rsun}{\hbox{$\hbox{R}_\odot$}}
\newcommand{\kms}{\hbox{${\rm km}\:{\rm s}^{-1}$}}

\begin{document}


\title{XTE J1118+480: A Metal-Rich Black \\ Hole Binary in the Galactic Halo.}


\author{Jonay I. Gonz\'alez Hern\'andez\altaffilmark{1,2,3}, Rafael
Rebolo\altaffilmark{1,4}, Garik Israelian\altaffilmark{1}, Emilios T.
Harlaftis\altaffilmark{5,6}, Alexei V. Filippenko\altaffilmark{7} and
Ryan Chornock\altaffilmark{7}} 


\altaffiltext{1}{Instituto de Astrof{\'\i }sica de Canarias, E-38205 La Laguna,
Tenerife, Spain: jonay@iac.es, rrl@iac.es, gil@iac.es}
\altaffiltext{2}{CIFIST Marie Curie Excellence Team}
\altaffiltext{3}{Observatoire de Paris-Meudon, GEPI, 5 place Jules Janssen, 92195 Meudon Cedex, France}
\altaffiltext{4}{Consejo Superior de Investigaciones Cient{\'\i }ficas, Spain}
\altaffiltext{5}{Institute of Space Applications and Remote Sensing, National
Observatory of Athens, PO Box 20048, Athens 118 10, Greece}
\altaffiltext{6}{In memoriam}
\altaffiltext{7}{Department of Astronomy, University of California, Berkeley, CA
94720-3411: alex@astro.berkeley.edu, chornock@astro.berkeley.edu}


\begin{abstract}
We present medium-resolution optical spectra of the secondary star 
in the high Galactic latitude black hole X-ray binary \mbox{XTE
J1118+480} and determine the abundance of Mg, Al, Ca, Fe, and
Ni in its atmosphere. For all the elements investigated we find
supersolar abundances; thus, we reject the hypothesis that the black
hole came from the direct collapse of an ancient massive halo star.
The compact primary most likely formed in a supernova event of a
massive star whose nucleosynthetic products polluted the secondary
star. The observed element abundances and their ratios can be
explained using a variety of supernova models with a wide range of
metallicities. While an explosive origin in the Galactic halo or thick
disk cannot be discarded, a metal-rich progenitor is clearly favored
by the observed abundance pattern. This suggests that the black hole
was produced in the Galactic thin disk with a violent natal kick,
propelling the X-ray binary to its current location and orbit.
\end{abstract}

\keywords{black hole physics --- stars: abundances --- stars: evolution --- 
stars: individual (\mbox{XTE J1118+480}) --- supernovae: general --- X-rays: 
binaries} 

\section{Introduction}

The low-mass X-ray binary \mbox{XTE J1118+480} was discovered with the
all-sky monitor aboard the Rossi X-ray Timing Explorer on UT 2000
March 29 (Remillard et al.\ 2000). Throughout its outburst the source
remained in the low/hard state, one of the characteristic spectral
states of an accreting black hole binary (McClintock et al.\ 2003).
The system consists of a black hole with a mass estimated in the range
6$-$8 {\Msun} and a late-type secondary star of 0.1$-$0.5 {\Msun}
(Wagner et al.\ 2001). 

The extraordinarily high Galactic latitude
($b \approx 62.3^{\circ}$), together with its distance of $1.85\pm0.36$ kpc
(Wagner et al.\ 2001), places the system at a height of $\sim$1.6 kpc 
above the Galactic plane. In addition, an accurate
measurement of its proper motion coupled with its distance provides
space-velocity components $U$, $V$ which seem consistent with 
those of some old halo globular clusters (Mirabel et al.\ 2001).
If the system formed in the Galactic halo, the black hole could be
either the remnant of a supernova in the very early Galaxy or the
result of a direct collapse of an ancient massive star. However,
the galactocentric orbit crossed the Galactic plane many times in the
past, and an alternative possibility is that the system formed in the
Galactic disk and was launched into its present orbit as 
a consequence of the ``kick'' acquired in the supernova explosion of a
massive star  (Gualandris et al.\ 2005). The metallicity of the secondary star may
provide a key to distinguish among these two possible birth places,
giving important clues to the formation of the black hole and the
properties of the supernova explosion, such as symmetry, released
energy, and characteristics of the ejected matter (Israelian et al.\
1999; Podsiadlowski et al.\ 2002).       

\section{Observations}

We obtained 74 medium-resolution spectra
($\lambda/\delta\lambda\approx 6,000$) of the secondary star in 
XTE J1118+480, in quiescence, on UT 14 February 2004,
using the 10-m Keck~II telescope, equipped with the Echellette
Spectrograph and Imager (ESI; Sheinis et al.\ 2002). The exposure time
was fixed at 300~s to minimize the effects of orbital smearing which,
for the orbital parameters of XTE J1118+480, is in the range 0.6--26.6
{\kms}, smaller than the instrumental resolution of $\sim 50$
{\kms}. Each individual spectrum was corrected for the radial velocity
of the star, and the spectra were combined in order to improve the
signal-to-noise ratio. After binning in wavelength in steps of 0.3
{\AA}, the final spectrum had an average signal-to-noise ratio of 80
in the continuum. The data cover the spectral range 4000--9000 {\AA}
and clearly show the characteristic emission lines of accreting
low-mass X-ray binaries (Balmer series and \ion{Ca}{2} near-infrared
triplet, \ion{He}{1} 5876 {\AA}, \ion{He}{1} 6678 {\AA}), superimposed 
on the typical photospheric spectrum of a late-type star.  

\section{Chemical Analysis}

A comparison of the observed spectrum with the spectra of ten template
stars (K0V--M2V) obtained with the same instrument 
allows us to classify the secondary as a mid to late K-type star. In
Figure 1 we display two spectral regions containing some relevant lines
for our analysis as well as a spectrum of a template star of similar
spectral type for comparison with available detailed chemical 
analysis (Allende Prieto et al.\ 2004). We also show synthetic line 
profiles for different stellar abundances computed with the local 
thermodynamic equilibrium (LTE) code MOOG (Sneden 1973), adopting the
atomic line data from the Vienna Atomic Line Database (VALD; Piskunov
1995) and using a grid of LTE model atmospheres (Kurucz 1993). 

In order to perform the chemical analysis of the secondary star
we used a technique which combines a grid of synthetic
spectra and a $\chi2$-minimization procedure that includes Monte-Carlo
simulations (Gonz\' alez Hern\'andez et al.\ 2004, 2005). First, we
inspected the observed spectrum in order to select the most suitable
features for a chemical abundance determination. We identified nine
spectral features containing in total 30 lines of \ion{Fe}{1} and 8
lines of \ion{Ca}{1} with excitation potentials between 1 eV and 5 eV.
The oscillator strengths of the re\-le\-vant spectral lines were
checked via spectral synthesis against the solar atlas (Kurucz et al.\
1984). We then generated a grid of about one million synthetic spectra
of each of these features, varying as free parameters the star effective
tem\-pe\-ra\-tu\-re ($T_{\mathrm{eff}}$), surface gravity ($\log g$),
and metallicity ([Fe/H]), together with the veiling from the accretion
disk which was assumed to be a linear function of wavelength, and thus
described by two additional parameters. Iron abundances were varied in
the range $-1.5 < \mathrm{[Fe/H]} < 1$ whereas the Ca abundance was
fixed, for each given iron abundance, according to the Galactic trend of
Ca (Bensby et al.\ 2005) for $\mathrm{[Fe/H]} < 0$, and fixed to
$\mathrm{[Ca/Fe]} = 0$ in the range $0 < \mathrm{[Fe/H]} < 1$. A
rotational broadening of 100 {\kms} and a limb darkening $\epsilon =
0.8$ were adopted. The microturbulence ($\xi$) was computed using an 
experimental expression as a function of effective temperature and
surface gravity (Allende Prieto et al.\ 2004). 
 
We compared, using a $\chi2$-minimization procedure, this grid with
1000 realizations of the observed spectrum. Using a bootstrap
Monte-Carlo method, we found the most likely values $T_{\mathrm{eff}}
= 4700 \pm 100$ K, $\log (g/{\rm cm~s}^2) = 4.6 \pm 0.3$,
$\mathrm{[Fe/H]} = 0.2 \pm 0.2$, and a disk veiling (defined as
$F_{\rm  disk}/F_{\rm total}$) of less than 40\% at 5000 {{\AA}} and
decreasing toward longer wavelengths. The (1$\sigma$) uncertainty in
the iron-abundance determination takes into account the uncertainties
in the stellar and veiling parameters. The effective temperature and
surface gravity are consistent with previous spectral classifications
and similarly the reported veiling values (Torres et al.\ 2004). Using
the derived stellar and veiling parameters, we analyzed several
spectral regions where we had identified various lines of Fe, Ca, Al,
Mg and Ni. Abundances of all the elements are listed in Table 1. The
1$\sigma$ uncertainty in the abundance determination takes into
account uncertainties in the stellar and veiling parameters.

Remarkably, we find a metallicity higher than solar, which is extremely
atypical of halo stars (Allende Prieto et al.\ 2006). 
In Figure 1 we show the best-fit synthetic
spectrum to various features in two different spectral regions, and
for comparison, a model with twenty-five times lower metal content. Notice the
inadequacy of low-metallicity models to reproduce the observed
features, even in the extreme case that no veiling is considered
(adding veiling would make the discrepancy much worse). An iron
abundance of $\mathrm{[Fe/H]}=-1.2$ is more than 6$\sigma$ away from
the best-fit solution ($\mathrm{[Fe/H]}=0.18$), and hence very 
unlikely. We have also found
that abundances of Al, Ca, Mg, and Ni are higher than solar (see Table
1). In Figure 2, we show that the abundance ratios of these elements
with respect to iron are consistent with those of stars in the solar
neighborhood from Gilli et al. (2006). We have also determined an
upper limit to the Li abundance $\log
\epsilon(\mathrm{Li})_\mathrm{LTE} = \log
[N(\mathrm{Li})/N(\mathrm{H})]_\mathrm{LTE} + 12 \le 1.61 \pm 0.25$
using the Li 6708 {\AA} line. This value seems to be lower
that typical high Li abundances measured in other late-type secondary
stars in soft X-ray transients whose origin is still an open question
(Mart{\'\i}n et al. 1994).

\section{Discussion and Conclusions}

If we include the metallicity distribution of halo and thick-disk 
stars (Allende Prieto et al.\ 2006) and thin-disk stars (Allende
Prieto et al.\ 2004) in the equations based on kinematics for
establishing the relative likehoods of belonging to the halo, thick 
disk, or thin disk (Bensby et al.\ 2003), the probability that a star with 
the Galactic space velocity components of this system (Mirabel et al.\
2001; $U=-105\pm16$ {\kms}, $V=-98\pm16$ {\kms}, $W=-21\pm10$ {\kms})
and metallicity $\mathrm{[Fe/H]} = 0.18$ belongs to the Galactic halo
is less than 0.1\%. Moreover, the kinematics alone suggest thick-disk
rather than halo membership, although the high metallicity of the
secondary star favors thin-disk membership. If, however, the progenitor
was a massive star in the halo or thick disk, the high metallicity of 
the secondary rules out the hypothesis that the black hole
was formed by direct collapse of the massive star; instead, a
supernova explosion origin is strongly suggested.

Given its present orbital distance from the black hole ($a_{\rm c}
\approx 3$ {\Rsun}, Wagner et al.\ 2001), it is plausible
that the secondary star captured a significant fraction of the matter
ejected in a supernova explosion. The chemical composition of the
secondary may provide crucial information on nucleosynthesis
in the progenitor and the formation mechanism of the black hole.  
We consider two possible scenarios for the origin of the compact
object: either it formed in the Galactic halo or thick disk as a
result of the explosion of a metal-poor massive progenitor
which enriched the secondary star from the typical abundances of
halo/thick-disk stars up to the observed supersolar values, or
alternatively, it formed in the Galactic thin disk with a natal kick
imparted during the supernova explosion which propelled the binary
into its current orbit. 

In the first case, the similarity with the kinematics of halo and
thick-disk stars makes unnecessary a significant kick during the black
hole formation process. However, since the typical metallicities of
halo stars and thick-disk stars are significantly lower than solar, it
is required that the secondary captured enough matter from the ejecta
to reach the current abundances. A metal-poor $\sim$1 {\Msun}
secondary star, initially placed at an orbital distance of $\sim$6
{\Rsun} (after tidal circularization of the orbit), would need to
capture roughly 5--10\% of the matter ejected in a spherically  
symmetric core-collapse supernova explosion of a 16 {\Msun} helium
core to achieve the observed iron abundance. We have considered
supernova and hypernova models of metal-poor progenitors with
different masses, metallicities, mass cuts (i.e., the mass above which
the matter is expelled at the time of the supernova explosion),
fallback (i.e., amount of mass which is eventually accreted by the
compact core), and mixing (Umeda \& Nomoto, 2002, 2005; Tominaga,
Umeda \& Nomoto 2006, in preparation), and assumed that all the
fallback material is well mixed with the ejecta. We find that a
sufficiently large Fe enrichment is possible for mass cuts in the
range 2--4 {\Msun} [and explosion energies $(1-30)\times10^{51}$
ergs]. The abundance ratios of the other elements are only marginally
reproduced by these models. An origin of \mbox{XTE J1118+480} in the
Galactic halo or in the Galactic thick disk cannot be discarded by the
present observations, nor can it be confirmed. Further higher-quality
observations and extensive model calculations are required to fully
explore this possibility. Further details of analysis will put
together in Gon\'alez Hern\'andez et al. (2006, in preparation).

In the second scenario, the system had to acquire a peculiar space
velocity of $\sim 180$ {\kms}, to change from a Galactic thin disk
orbit to the currently observed orbit (Gualandris et al.\ 2005), requiring
an asymmetric kick. Such kicks imparted during the birth of nascent neutron
stars, due to asymmetric mass ejection and/or an asymmetry in the
neutrino emission (Lai et al.\ 2001), have been proposed to
explain the large transverse motions of neutron stars in the plane of
the sky (Lyne \& Lorimer 1994). The black hole could have formed 
in a two-stage process where the initial collapse led to the
formation of a neutron star accompanied by a substantial kick and the
final mass of the compact remnant was achieved by matter that fell
back after the initial collapse as proposed to explain the origin of
the black hole in \mbox{Nova Sco 1994} (Podsiadloski et al.\ 2002;
Brandt et al.\ 1995).  

Spherically symmetric explosion models are able to explain the
observed metal enrichment. A modest amount of ejecta ($\sim 0.02$)
{\Msun} captured by a secondary with initial solar abundance is 
sufficient if vigorous mixing (Kifonidis et al.\ 2000)
between the fallback matter (which is necessarily large given the mass
of the black hole) and the ejecta took place. In this scenario the
black hole would be formed in a mild explosion with fallback as in
collapsar models (MacFadyen et al.\ 2001) associated with rapidly rotating
massive stars. However, in the pure spherical supernova explosion
scenario for \mbox{XTE J1118+480}, the maximum allowed ejected mass to
keep the secondary star gravitationally bound is $\Delta M \approx 8$
{\Msun}. This leads to a system velocity of $\sim$60 {\kms} (there
could be extreme cases with system velocities as high as $\sim 100$
{\kms} requiring very \emph{special} model parameters such as higher
secondary and black hole masses just after the explosion; see further
details in Sec. 3.2 of Gualandris et al.\ 2005), much lower
than the observed value, and a neutrino-induced kick is required 
if we are to explain the kinematics of the system.

The chemical composition of the ejecta in a non-spherically symmetric
supernova explosion is strongly dependent on direction. In particular,
if we assume that the jet is collimated perpendicular to the orbital
plane of the binary, where the secondary star is located, elements
such as Ti, Ni, and Fe are mainly ejected in the jet direction, while
Al, O, Si, S, and Mg are preferentially ejected near the equatorial
plane of the helium star (Maeda et al.\ 2002). Using predictions
for an aspherical explosion model of a 16 {\Msun} He core metal-rich
progenitor, we can  explain the observed abundances in the secondary.
Complete lateral mixing (Podsiadlowski et al.\ 2002) is required to account
for the similar enhancement of Mg and Ni and the observed abundances
can be reproduced for all mass cuts in the range 2--8 {\Msun}. 

It is therefore plausible that the black hole in \mbox{XTE
J1118+480} formed in the Galactic thin disk from a massive
metal-rich progenitor and was launched into its current orbit either
by mass ejection in an asymmetric supernova/hypernova explosion or by
a neutrino-induced kick. In addition, ultraviolet observations of the
accretion disk in this system suggest that the material accreted onto the
compact object is substantially CNO processed (Haswell et al.\ 2002),
indicating that the zero-age mass of the secondary star could have
been $\sim 1.5$ {\Msun}. By means of binary evolution calculations
this may constrain the age of the system in the range 2--5.5 Gyr
(Gualandris et al.\ 2005). Future chemical studies of the secondary
star during periods of quiescence may provide accurate abundances of 
elements (e.g., C, Ti, Si) whose lines either are not available in the
observed spectrum or are present in spectral regions where the
signal-to-noise ratio is too low for an accurate chemical analysis.
These may reveal further details of the formation mechanism of the
black hole in this system.  

\acknowledgments

We are grateful to Hideyuki Umeda, Ken'ichi Nomoto, and Nozomu Tominaga
for sending us their explosion models for metal-poor and metal-rich
progenitors and several programs for our model computations. We also
thank Keiichi Maeda for providing us with his aspherical explosion
models, and for helpful discussions. We are grateful to Tom Marsh for
the use of the MOLLY analysis package, to Jorge Casares for his helpful
comments on different aspects of this work, and to Ryan J. Foley for
assistance with the observations. We also thank the referee for
helpful comments. The W. M. Keck Observatory is
operated as a scientific partnership among the California Institute of
Technology, the University of California, and NASA; it was made
possible by the financial support of the W. M. Keck Foundation. This
work has made use of the VALD database and IRAF facilities. It was
funded in part by Spanish Ministry project AYA2005-05149 and by
US National Science Foundation grant AST-0307894. We dedicate this
paper to the memory of our dear friend and collaborator E. T.
Harlaftis, whose life was tragically cut short by a snow avalanche on
13 February 2005.

\clearpage

\begin{figure}
\epsscale{0.60}
\plotone{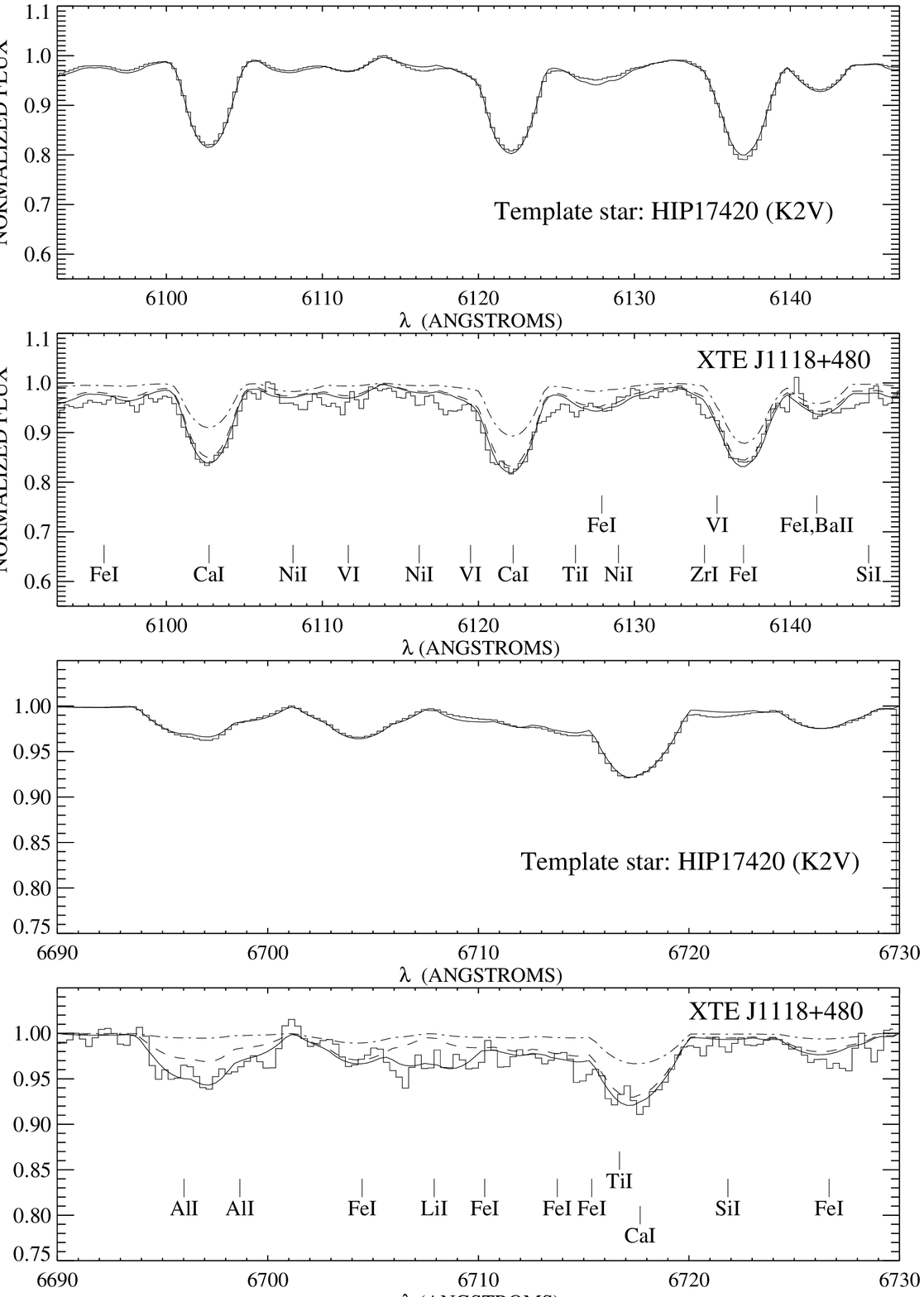}
\caption{Best synthetic spectral fits to the ESI
spectrum of the secondary star in the \mbox{XTE J1118+480} system
(second and bottom panels) and the same for a template star
(properly broadened) taken from Allende Prieto et al. (2004)
shown for comparison (top and third panels). Synthetic spectra are
computed for typical abundances for a halo star ($\mathrm{[Fe/H]}=-1.2$,
dashed-dotted blue line), solar abundances ($\mathrm{[Fe/H]}=0$,
dashed green line), and best-fit abundances (solid red line). In
addition, note that for solar and best-fit abundances we have
applied to the synthetic spectra the corresponding values for the
veiling according to the solution found with the fitting procedure.
However, for the low-metallicity synthetic spectra, we have not
assumed any veiling.}   
\label{fig1}
\end{figure}

\clearpage

\begin{figure}
\epsscale{0.80}
\plotone{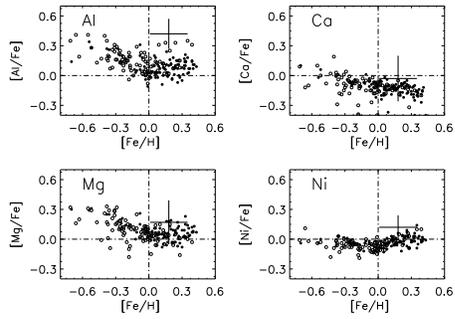}
\caption{Abundance ratios of the secondary star 
in XTE J1118+480 (blue wide cross) in comparison with the
abundances of G and K metal-rich dwarf stars. Galactic trends were
taken from Gilli et al. (2006). The size of the cross 
indicates the uncertainty. Filled and empty circles correspond to
abundances for planet host stars and stars without known planet
companions, respectively. The dashed-dotted lines indicate solar
abundance values.} 
\label{fig2}
\end{figure}

\clearpage

\begin{deluxetable}{lccccc}
\tablewidth{7cm}
\tabletypesize{\scriptsize}
\tablecaption{Abundances in the secondary star}   
\tablehead{\colhead{Parameter} & \colhead{Al}& \colhead{Ca}&
\colhead{Mg}& \colhead{Fe}& \colhead{Ni}}
\startdata
{$\mathrm{[X/H]}$}	  & 0.60 & 0.15 & 0.35 & 0.18 & 0.30 \\
{$\Delta \mathrm{[X/H]}$} & 0.20 & 0.23 & 0.25 & 0.17 & 0.21 \\
\enddata
\tablecomments{\scriptsize{Element abundances (calculated assuming LTE) are
$\mathrm{[X/H]}= \log [N(\mathrm{X})/N(\mathrm{H})]_{\rm star} - \log
[N(\mathrm{X})/N(\mathrm{H})]_{\rm Sun}$, where $N(\mathrm{X})$ is
the number density of atoms. Uncertainties, $\Delta \mathrm{[X/H]}$, are
1$\sigma$ and take into account uncertainties in the
stellar and veiling parameters.}} 
\end{deluxetable}

\end{document}